# Impact of pulse dynamics on timing jitter in mode-locked fiber lasers

Youjian Song, Kwangyun Jung, and Jungwon Kim[*]

KAIST Institute for Optical Science and Technology and School of Mechanical, Aerospace and Systems Engineering,
Korea Advanced Institute of Science and Technology (KAIST), Daejeon 305-701, Korea
*Corresponding author: jungwon.kim@kaist.ac.kr

We investigate the high-frequency timing jitter spectral density of mode-locked fiber lasers in different mode-locked regimes. Quantum noise-limited timing jitter spectra of mode-locked-regime-switchable Yb fiber lasers are measured up to the Nyquist frequency with sub-100 as resolution. The integrated rms timing jitter of soliton, stretched-pulse, and self-similar Yb fiber lasers is measured to be 1.8 fs, 1.1 fs, and 2.9 fs, respectively, when integrated from 10 kHz to 40 MHz. The distinct behavior of jitter spectral density related to pulse formation mechanisms is revealed experimentally for the first time. © 2011 Optical Society of America

Femtosecond mode-locked fiber lasers have recently attracted great attention for many high-precision time-frequency applications [1,2] such as optical frequency combs, low-noise microwave signal synthesis, photonic analog-to-digital conversion, arbitrary optical waveform generation, and large-scale timing distribution. Further advances in this area require detailed study on noise properties of mode-locked fiber lasers. In particular, accurate characterization of timing jitter is important for the optimization of mode-locked fiber lasers toward lower timing jitter and phase noise performance. There have been several theoretical and numerical studies on timing jitter of mode-locked fiber lasers in the context of intra-cavity dispersion and filtering effects [3-5]. However, due to inherent low jitter approaching sub-fs level, conventional measurement methods based on photo-detection and microwave mixers could not measure the jitter spectral density up to the full Nyquist frequency [6]. Balanced optical cross-correlation (BOC) [7] has emerged as a powerful tool for high dynamic range timing jitter characterization of optical pulse trains with sub-fs resolution, and has been successfully used for timing jitter measurement of Er fiber lasers [8] and Cr:LiSAF solid-state lasers [9]. Such sub-fs-resolution measurement capabilities now make it possible to observe the impact of intracavity pulse dynamics on timing jitter over the full Nyquist frequency.

In this Letter, we characterized the mode-locked regime dependent, quantum-limited timing jitter spectra of fiber lasers over the Nyquist frequency for the first time. We characterized the jitter spectral density of three typical mode-locked regimes of Yb fiber lasers – soliton, stretched-pulse, and self-similar [10]. We could observe clear difference in the shape and relative magnitude in measured jitter spectral density between different mode-locked regimes. The measured integrated rms timing jitter is 1.8 fs, 1.1 fs, and 2.9 fs [when integrated from 10 kHz to 40 MHz offset frequency] for soliton, stretched-pulse, and self-similar regimes, respectively.

Two home-built, nearly identical Yb fiber lasers are used to study the mode-locked condition dependent timing jitter. A standard nonlinear polarization evolution (NPE) Yb fiber laser design is employed [11]: one laser is based on a unidirectional cavity, and the other one is an σ-cavity with a piezoelectric transducer (PZT)-mounted mirror for repetition rate locking. The gain medium is a segment of 23 cm Yb-doped fiber with 1200 dB/m absorption at 976 nm. The total fiber length is 2 m, and the repetition rate is set to 80 MHz. A 600 line/mm grating pair is employed for intracavity dispersion compensation. We can operate the Yb fiber laser in different mode-locked regimes by adjusting the amount of dispersion using the grating pair and finding the right NPE condition using waveplates. The intracavity dispersion is set to the typical conditions for each mode-locked regimes: large negative dispersion for soliton and slightly positive dispersion for stretched-pulse and self-similar. Specifically, intracavity dispersion of −0.021 $ps^2$, +0.003 $ps^2$, and +0.006 $ps^2$ at 1030 nm center wavelength is used to study the timing jitter characteristics of soliton, stretched-pulse, and self-similar conditions, respectively. The corresponding optical spectra are shown in the inset of Fig. 2. As the output coupler is located at the maximum chirp position in the cavity, output pulses are dechirped by extracavity prism pair (for soliton lasers) or grating pair (for stretched-pulse and self-similar lasers) to maximize timing resolution of the BOC. The laser output parameters are summarized in Table 1.

Table 1. Summary of experiment parameters

| Regime | SOL | SP | SS |
|---|---|---|---|
| Laser output | | | |
| Pulse width (fs) | 350 (180) | 740 (57) | 1800 (69) |
| Average power (mW) | 40 (32) | 75 (30) | 100 (40) |
| BOC-based timing jitter measurement | | | |
| Detection sensitivity (V/fs) | 0.0047 | 0.0136 | 0.0178 |
| Sum-frequency power (mW) | 0.14 | 0.25 | 0.28 |
| Shot noise level ($fs^2$/Hz) | $1.8\times10^{-10}$ | $1.4\times10^{-11}$ | $8\times10^{-12}$ |
| Resolution[*] (as) | 84 | 23.7 | 18 |

SOL: soliton; SP: stretched-pulse; SS: self-similar.
Values in parenthesis measured after dispersion compensation.
[*] Integration in 40 MHz bandwidth.

The timing jitter of Yb fiber lasers is characterized by the BOC method. The experimental setup is shown in Fig. 1. A 400 μm thick type-II phase-matched BBO [9] optimized for 1030 nm is used for both generating sum-frequency signals and providing group delay between two output pulse trains from the two lasers. Another piece of birefringence glass with different thickness (BG in Fig. 1) is employed in the BOC to provide additional group delay for different pulse durations of three mode-locked regimes. The sum-frequency signal from the BBO is transmitted by a dichroic mirror (DM2 in Fig. 1) and detected by one port of a Si balanced photodetector. The group delayed fundamental light pulses are reflected by the same mirror and focused into the BBO again. The backward-generated sum-frequency signal is reflected by a second dichroic mirror (DM1 in Fig. 1) and collected by the other port of the balanced photodetector. The subtracted output from balanced detector is proportional to the timing offset between the two input pulses [the BOC output is shown in the inset of Fig. 1]. In order to confine the measurement in the linear detection range of BOC, the two lasers are synchronized by a low-bandwidth (<7 kHz in this work) phase-locked loop. An rf spectrum analyzer is used to measure the cross-correlation signal outside the locking bandwidth, which follows the sum of timing jitter power spectral density of two lasers. The measured spectral density is divided by two to get the spectral density of a single laser since the two lasers are nearly identical and uncorrelated. Table 1 summarizes the BOC performances for three different mode-locked conditions.

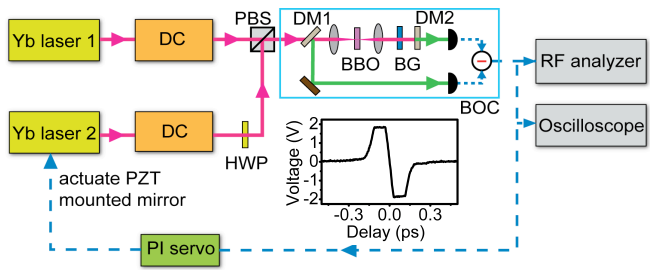

Fig. 1. (Color online) Experimental setup of Yb fiber laser timing jitter measurement based on BOC. BG, birefringence glass plate; DC, dispersion compensation; DM, dichroic mirror; HWP, half-wave plate; PBS, polarization beamsplitter. The inset shows the measured cross-correlation trace of two lasers without synchronization. Solid and dashed lines indicate optical and electric signal paths, respectively.

The measured timing jitter power spectral density and the corresponding integrated jitter for the three mode-locked regimes are shown in Fig. 2. The measurement resolution of the three mode-locked conditions is different because of the different output pulse width and power. As listed in Table 1, self-similar and stretch-pulse lasers have sub-100 fs extracavity dechirped pulse duration, which guarantees that the timing jitter measurement is not limited by shot noise level over the entire Nyquist frequency. For soliton mode locking, due to the limited resolution of long pulse width (180 fs after dechirping), the measurement is restricted by shot noise above 8 MHz, as shown by dashed line of Fig. 2. The resolution-limited part of jitter spectrum above 8 MHz contributes ~100 as to the total integrated jitter.

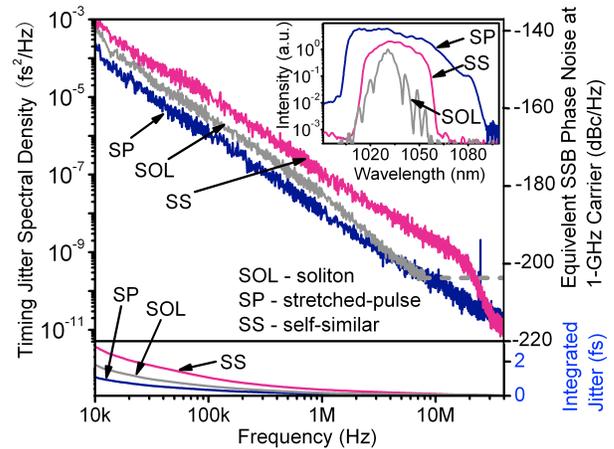

Fig. 2. (Color online) Timing jitter measurement results. Top: The measured timing jitter power spectral density and the equivalent single-sideband (SSB) phase noise at 1-GHz carrier frequency. Bottom: The integrated rms timing jitter. Inset: The optical spectra of laser outputs.

The amplified spontaneous emission (ASE) quantum noise can simultaneously introduce timing jitter directly in the time domain (directly-coupled jitter) and indirectly through the optical frequency domain (indirectly-coupled jitter) [3]. In the former case, due to the white noise characteristic of the ASE noise, the pulse position undergoes a random walk, which results in timing jitter spectral density of $1/f^2$ slope. In Fig. 2 the stretched-pulse laser measurement result clearly shows such a $1/f^2$ slope over the entire offset frequency range, which indicates the quantum-limited random walk nature directly originated from the ASE noise. The ASE noise can also introduce additional timing jitter by center frequency fluctuations coupled via intracavity dispersion. The timing jitter spectral density caused by this indirect ASE noise coupling results in $1/f^2$ slope at low offset frequency range and $1/f^4$ slope at high offset frequency range [5]. The soliton and self-similar laser measurement results in Fig. 2 show the combined effects of both directly- and indirectly-coupled timing jitter originated from the ASE noise. The indirectly-coupled timing jitter is dominant in the low offset frequency, whereas only the directly-coupled timing jitter is remained in the high offset frequency. As a result, there is a transition frequency region which features a jitter spectrum slope between $1/f^4$ and $1/f^2$, depending on the difference in directly- and indirectly-coupled timing jitter levels. The self-similar laser has much higher indirectly-coupled jitter than the soliton laser, and the transition from indirectly-coupled jitter to directly-coupled jitter with $1/f^4$ slope is clearly visible in the 20 MHz – 30 MHz range. These quantum-limited timing jitter spectra were previously predicted by theory and numerical simulations [5], and this is the first time to observe them experimentally owing to the attosecond-resolution BOC measurement technique.

The measured integrated rms timing jitter is 1.8 fs, 1.1 fs, and 2.9 fs for soliton, stretched-pulse and self-similar laser conditions, respectively, when integrated from 10 kHz to 40 MHz (Nyquist frequency). Note that the measured timing jitter of stretched-pulse regime ($10^{-4}$ fs$^2$/Hz at 10 kHz offset frequency and 1.1 fs integrated jitter from 10 kHz to 40 MHz) is, to our knowledge, the lowest high-frequency timing jitter from passively mode-locked fiber lasers so far. The measurement results in Fig. 2 also show that even the timing jitter spectrum shape strongly depends on the mode-locked regimes and intracavity pulse dynamics, the integrated jitter numbers are all in the similar level of 1 – 3 fs range.

Recent study on carrier-envelope offset frequency ($f_{ceo}$) noise in mode-locked Yb fiber lasers showed that intracavity dispersion is the dominant factor for frequency noise [12]. As it is known that timing jitter and $f_{ceo}$ noise are closely related [5], we performed additional measurement to investigate whether the intracaivty dispersion is the dominant factor for timing jitter as well. For this, we set the intracavity dispersion at a moderately positive value (+0.004 ps$^2$) where the Yb fiber laser can be operated in either stretched-pulse or self-similar regimes. By rotating the intracavity waveplates, the NPE strength changes and subsequently the mode-locked regime can be switched between stretched-pulse and self-similar. The intracavity pulse energy in both mode-locked regimes is set to equal by adjusting the pump power. Then the major difference between the two regimes is the round-trip pulse dynamics, especially the average pulse width and pulse chirp. Fig. 3 shows the measured timing jitter spectra of the stretched-pulse and self-similar regimes with the same cavity dispersion and pulse energy. Interestingly, even at the same intracavity dispersion, the timing jitter spectrum is significantly different between the two regimes, particularly the indirectly-coupled timing jitter part in the low offset frequency. This measurement result suggests that the timing jitter characteristic indeed strongly depends on the intracavity pulse dynamics set by the mode-locked regime, rather than the intracavity dispersion alone.

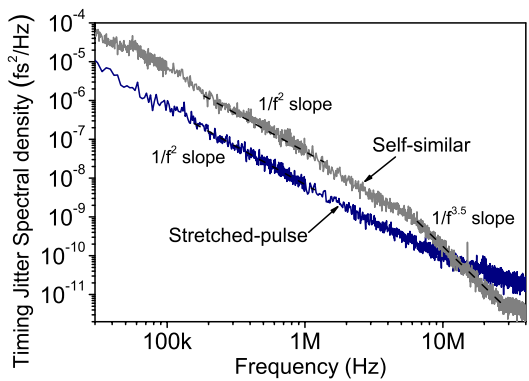

Fig. 3. (Color online) Comparison of timing jitter power spectral density of self-similar and stretched-pulse regimes at the same intracavity dispersion condition (+0.004 ps$^2$) and pulse energy.

For the optimization of timing jitter performance, a stretched-pulse laser can be used because it generates much shorter pulses with higher pulse energy than a soliton laser. However, timing jitter reduction in the stretched-pulse regime is not straightforward because of its large chirp parameter at typically used slightly positive dispersion condition [4,9]. As a result, as shown in Fig. 2, although the magnitude of dispersion is much larger for soliton lasers, the amount of jitter is not much different from that of stretched-pulse lasers. Thus, further reduction of timing jitter requires the operation of stretched-pulse lasers at a different dispersion condition. It was theoretically predicted that slightly negative dispersion can lead to the lowest timing jitter by combining the shortest pulse width and much reduced chirp parameter [4]. Future work will be focused on the exploration of jitter characteristics in the slightly negative to close-to-zero dispersion range for the optimization of timing jitter in mode-locked fiber lasers.

In summary, the ASE quantum noise-limited timing jitter in Yb fiber lasers is studied over the full Nyquist frequency for the first time using an attosecond-resolution BOC. The measured integrated rms timing jitter is 1.8 fs, 1.1 fs, and 2.9 fs [when integrated from 10 kHz to 40 MHz offset frequency] for soliton, stretched-pulse, and self-similar regimes, respectively. We further experimentally showed that the pulse formation mechanism strongly influences the shape and relative magnitude in jitter spectral density between different mode-locked regimes.

The authors would like to thank Jonathan A. Cox for discussions on laser locking and Hyoji Kim and Chur Kim for technical support. This research was supported in part by Pohang Accelerator Laboratory and Basic Science Research Program through the National Research Foundation of Korea (NRF) funded by the Ministry of Education, Science and Technology (2010-0003974).